\documentclass[aps, preprint, 11pt, amsmath,amssymb,floatfix,graphicx,pre]{revtex4-2}

\draft
\usepackage{graphicx}
\usepackage{dcolumn}
\usepackage{bm}
\usepackage[mathlines]{lineno}
\usepackage[english]{babel}
\usepackage{setspace}
\usepackage{color}
\usepackage{float}
\usepackage[utf8]{inputenc}
\usepackage[T1]{fontenc}
\usepackage{mathptmx}
\DeclareMathAlphabet{\mathcal}{OMS}{cmsy}{m}{n}
\usepackage{etoolbox}

\begin{document}

\title[]{A generalized theory of Brownian particle diffusion in shear flows}
\author{Nan Wang}
\affiliation{Faculty of Aerospace Engineering, Technion - Israel Institute of Technology, Haifa, 320003, Israel.}
\author{Yuval Dagan}
\email{yuvalda@technion.ac.il}
\affiliation{Faculty of Aerospace Engineering, Technion - Israel Institute of Technology, Haifa, 320003, Israel.}

\begin{abstract}
This study presents a generalized theory for the diffusion of Brownian particles in shear flows.
By solving the Langevin equations using stochastic instead of classical calculus, we propose a new mathematical formulation that resolves the particle MSD at all time scales for any two-dimensional parallel shear flow described by a polynomial velocity profile. 
We show that at long-time scales, the polynomial order of time in the particle MSD is $n+2$, where $n$ is the polynomial order of the transverse coordinate of the velocity profile. 
We generalize the theory to resolve particle diffusion in any polynomial shear flow at all time scales, including the order of particle relaxation time scale, which is unresolved in current theories.
Particle diffusion at all time scales is then studied for the cases of Couette and plane-Poiseuille flows and a polynomial expansion of a hyperbolic tangent flow while neglecting the boundary effects.
We observe three main stages of particle diffusion along the timeline for which the particle MSD is distinctly different due to different dominated physical mechanisms.
Thus, higher temporal and spatial resolution for diffusion processes in shear flows may be realized, suggesting a more accurate analytical approach for the diffusion of Brownian particles.
\end{abstract}

\maketitle

\newpage
\section{Introduction}
The diffusion of Brownian particles suspended in shear flows has been studied comprehensively in science and engineering. Examples can be found in aerospace propulsion \cite{hsu2018review}, filtration of aerosols \cite{tcharkhtchi2021overview}, atmospheric flows \cite{greenfield1957rain, ardon2015laboratory}, medical applications \cite{darquenne2020deposition}, and biological studies~\cite{huber2019brownian}. 
Settling of submicron particles that may be affected by Brownian diffusion is also of particular interest in transmission routes of viral diseases, where complex particle-flow interactions occur \cite{tellier2019recognition, feng2020influence, dagan2021settling, avni2022dynamics, avni2022dispersion}, and was shown to be particularly sensitive in vortical shear flows~\cite{dagan2017particle, dagan2017similarity, avni2023droplet1, avni2023droplet2}.
In the present study, we shall derive a theoretical method to address the influence of shear flows on particle diffusion in a generalized framework.

Brownian motion was first observed by the botanist Robert Brown in 1827~\cite{brown1828xxvii} and was studied mathematically by Einstein in 1905~\cite{einstein1956investigations, einstein1905motion} and Smoluchowski in 1906~\cite{smoluchowski1906kinetic}. They obtained the particle mean squared displacement (MSD) in a quiescent medium as
$ \langle s_p^2\rangle = 2 D t \label{eins}$, where $D$ is the diffusion coefficient and $t$ is time. This relation was verified experimentally by Jean Perrin et al. in 1908~\cite{brush1968brownian, genthon2020concept}.
Langevin~\cite{lemons1997paul} introduced a random force into Newton's second law and obtained the Einstein-Smoluchowski formula through the following equation
 \begin{equation}
    m \ddot{x}_p(t)=- \gamma\dot{x}_p(t)+ N(t)  ~,           \label{langevini}
 \end{equation}
where $m$ is the particle mass, $x_p(t)$ is the particle position, $\gamma \dot{x}_p(t)$ is the drag force, and $N(t)$ is a time-dependent random force.
Uhlenbeck and Ornstein \cite{uhlenbeck1930theory} analytically solved the Langevin equation for a free particle suspended in a stationary medium and derived the probability distribution function for the particle velocity and displacement. In their analysis, the particle velocity distribution coincides with the one-dimensional Maxwell–Boltzmann distribution. It has been recently verified by conducting experiments of particles in gaseous \cite{li2010measurement} and liquid \cite{kheifets2014observation, mo2015testing} mediums using optical tweezers. Their solution of particle displacement follows a Gaussian process, and the particle MSD is asymptotic to
\begin{equation}
\langle s_p^2 \rangle  \rightarrow  \left  \{ \begin{array}{ll} u_p^2(0) t^2  & \ \ t \to 0 \\  2 D t  & \ \ t \to \infty  \end{array} \right. ,
\end{equation} 
which has also been recently confirmed by experiments \cite{li2010measurement, li2013brownian,donado2017brownian} for both short-time and long-time scales.

Inspired by Perrin's description of Brownian trajectories, Wiener \cite{winer1966mathematician} worked on the properties of Brownian trajectories and characterized Brownian motion as continuous non-differentiable curves, which led to a new field of study on stochastic processes. Hence, a disagreement about the existence of the velocity of the Brownian particle between Wiener's theory and Ornstein-Uhlenbeck's theories emerged.
In 1942, Doob \cite{doob1942brownian} devised a theory of stochastic processes with continuous parameters and formally presented the Langevin equation in a differential manner,
\begin{equation}
\left\{ \begin{array}{l}  d u_p  = -\frac{\gamma}{m} u_p(t) d t +  d \mathbb{B}(t)  \\  d x_p(t) = u_p(t) d t \end{array} \right.  ~. \label{doob}
\end{equation}  
He showed that $\mathbb{B}(t)$ satisfies the properties of a Wiener process and, therefore, circumvented the argument on the differentiability of Brownian trajectories as a consequence of expressing the Langevin equation in a differential form instead of using derivatives. In the present study, we shall use the stochastic formula of the Brownian force of equation~(\ref{doob}).

As for Brownian motion in a medium with external forces, Uhlenbeck et al. \cite{uhlenbeck1930theory} studied the dynamics of Brownian particles in harmonic flows by solving Langevin equations with an additional term. 
Furthermore, particle diffusion in shear flows was also investigated analytically and experimentally~\cite{van1977diffusion,foister1980diffusion}. In an unbounded linear shear flow, the diffusion of Brownian particles was first derived by Foister et al. \cite{foister1980diffusion, van1977diffusion} and more recently by Katayama et al.~\cite{katayama1996brownian} and Chakraborty~\cite{chakraborty2012time}. They obtained the particle MSD along the streamlines at a time much longer than the particle relaxation time ($\tau_p = m/\gamma$) as
\begin{equation}
   \langle s_p^2\rangle = 2 D t \Big (1+\frac{1}{3}\alpha^2 t^2 \Big )~,
\end{equation}
where $\alpha$ is the shear rate \cite{foister1980diffusion, van1977diffusion}. Particle diffusion is enhanced compared to the Einstein-Smoluchowski diffusion equation and was coined \textit{anomalous diffusion} due to the rightmost term, which is proportional to $t^3$. 
Recently, Takikawa and Orihara \cite{takikawa2012diffusion, takikawa2019three, orihara2011brownian} performed experiments with a stereo microscope to validate the $t^3$-term and the dependence on the shear rate $\alpha$. Their results verified the anomalous diffusion in the streamwise direction and the correlation between the particle MSD in the streamwise direction and the velocity gradient, which is $\alpha$ in this case. Takikawa et al. \cite{takikawa2012diffusion} also investigated the diffusion of Brownian particles under a sinusoidal oscillatory shear flow. The results in their experiments are in good agreement with the theoretical results obtained from solving the Langevin equation. This study also confirms that the diffusion in the streamwise direction is related to the velocity variation in the transverse direction.

Particle dispersion in the streamwise direction of an unbounded Poiseuille flow was derived by Taylor \cite{taylor1953dispersion, frankel1989foundations} and generalized by Aris \cite{aris1956dispersion} through the convection-diffusion equation. Taylor first proposed the coupling between the Brownian diffusion in the transverse direction and the velocity gradient of the flow, which has been confirmed experimentally and theoretically.

Furthermore, Foister et al. obtained a $t^4$ diffusion term of the particle MSD in the streamwise direction for unbounded plane-Poiseuille and Hagen-Poiseuille flows, which was qualitatively verified by experiments \cite{foister1980diffusion}. By solving the Langevin equation with an additional term, they obtained the particle MSD in an unbounded plane-Poiseuille flow at a time much longer than $\tau_p$ as
\begin{equation}
    \langle s_p^2\rangle = 2 D t \Big (1+\frac{4}{3}\zeta^2 y_0^2 t^2 + \frac{7}{6}D\zeta^2 t^3 \Big )~,
\end{equation}
where $\zeta = V_{max}/R^2$, $V_{max}$ is the maximum velocity at the centerline of the flow, $R$ is half the distance between the two plates, and $y_0$ is the initial position of the particle in the transverse direction.
Foister's studies indicate that Langevin's approach to random movement provides another way to comprehend the effects of the coupling between particle diffusion and flow variation in the transverse direction; the $t^3$ and $t^4$ diffusion terms reveal that this coupling significantly alters the diffusion in the streamwise direction.

The review above indicates that the flow velocity gradient significantly affects the Brownian particle diffusion in shear flows.
Here, we may separate the diffusion process into three distinct temporal regions: A short-time scale much smaller than the particle relaxation time $\tau_p$, an intermediate time scale in the order of $\tau_p$, and a long-timescale much longer than $\tau_p$. As aforementioned, the particle MSD in unbounded linear shear flow and Poiseuille flow at long-time scales has been studied analytically and experimentally. 
The particle MSD for short time scales has not been studied analytically for different shear flows. Nevertheless, one would expect it to be the same as obtained by Uhlenbeck and Ornstein for stationary medium.
However, the particle MSD at intermediate time scales has yet to be discussed.

The particle diffusion in polynomial shear flows is ideal for studying anomalous diffusion since the velocity profile of a two-dimensional parallel laminar flow can be expanded by polynomial series.
However, the study for Brownian particle diffusion in parallel shear flows was so far limited to Couette and Poiseuille flows, of which the velocity profile is linear and parabolic functions of transverse coordinate, respectively. Moreover, current analytical solutions are restricted to the limit of either short or long time scales.
Thus, the present research aims to analytically derive the diffusion of Brownian particles in polynomial parallel laminar flows over all time scales by solving the Langevin equation using stochastic calculus.

In section \ref{mathSEC}, the Langevin equation for a Brownian particle diffusion and the deduction for the stochastic formula of the random force are presented. 
In the next section, we derive particle dynamics in a general polynomial parallel shear flow. 
The long-timescale asymptotics for the general polynomial flow is derived in section \ref{HighOrderLF} and is validated with the results of Foister et al. \cite{foister1980diffusion} for unbounded Couette flow and plane-Poiseuille flow at long-time scales. 
The particle MSD in unbounded Couette flow and plane-Poiseuille flow for all time scales are then presented in sections \ref{CouetteF} and \ref{PPoiseF}. 
In section \ref{PHTF}, we present a new solution employing the stochastic method and resolve the Brownian particle diffusion in a shear flow described by a hyperbolic tangent profile by expanding the velocity into polynomial series as an example. 
Finally, we discuss the observations in section \ref{disc}. 

\section{Mathematical Model}   \label{mathSEC}

Submicron-sized particles suspended in fluids may be subjected to drag forces, electrostatic forces, gravitational and other body forces, and the random force resulting from frequent collisions by surrounding molecules. The dynamics of a Brownian particle suspended in a flow can be described by the following Langevin equation
\begin{equation}
\left\{ \begin{array}{l} \frac{d\textbf{X}_p(t)}{d t} = \textbf{U}_p(t) \\  \\ m\frac{d\textbf{U}_p(t)}{d t}=\gamma \Big ( \textbf{V}_f(t)-\textbf{U}_p(t)\Big ) + \textbf{F}_B + \textbf{N}(t)\end{array} \right. ~,        \label{langevine}
\end{equation}      
where $\textbf{X}_p$ is the particle position vector; $\textbf{F}_{B}$ represents other body forces, such as gravity, electric forces, etc.; $\textbf{N}(t)$ is the Brownian force vector with the components $N_i(t)$ where $i$ refers to spatial coordinates $x$, $y$ or $z$; $\gamma [\textbf{V}_f(t)-\textbf{U}_p(t)]$ is the drag force vector, which is proportional to the difference between the particle velocity vector $\textbf{U}_p(t)$ and the flow velocity vector $\textbf{V}_f(t)$. The drag coefficient may be written as,
\begin{equation}
 \gamma =\frac{3\pi \mu d_p}{C_c}  \label{gamma}~,
\end{equation}
where $\mu$ is the dynamic viscosity of the medium, $d_p$ is the diameter of the particle, and $C_c$ is the Cunningham correction coefficient to the Stokes drag. 

The random force $\textbf{N}(t)$ results from frequent random collisions, which could be as often as $10^{20}$ times per second, between the particle and surrounding molecules in the medium. The extremely frequent collisions lead to memory loss at different time intervals.
Since random collisions are uniform in all directions, the average of the random force should be zero. Thus, by decomposing the random force into Cartesian coordinates, the random force in each direction $N_i(t)$ can be summarised,
by defining the mean and correlation of $N_i(t)$ at any time $t$ and $s$, as proposed by Uhlenbeck et al. \cite{uhlenbeck1930theory},
\begin{equation}
\langle N_i (t) \rangle =0 ;\ \  \ \   \langle N_i(t) N_j(s) \rangle = r \delta_{i,j}\delta(t-s)~, \label{assumption}
\end{equation}
where $\sqrt{r}$ is the magnitude of the random force, $\delta_{i,j}$ is the Kronecker delta function and $\delta$ is the Dirac delta function.
Suppose $t_B$ is the minimum time scale during which many collisions are occurring, such that these collisions eliminate all correlations between occurrences during $t_B$ and the preceding ones.
We shall denote
\begin{equation}
    d\mathbb{U}_i(t) = N_i(t) d t    \label{du=ndt}
\end{equation}
and investigate $\mathbb{U}_i(t)$ for time $t$ much longer than $t_B$. Dividing $t$ into intervals $0=t_0<t_1<t_2<\cdots<t_n=t$, where $t_k-t_{k-1}$ is in the order of $t_B$, then
\begin{equation}
    \mathbb{U}_i(t) - \mathbb{U}_i(0) = \sum_{k=1}^{n} [\mathbb{U}_i(t_k) - \mathbb{U}_i(t_{k-1})]~.
\end{equation}
Since during the time interval $t_k-t_{k-1}$ the random force $N_i(t)$ is independent of that before $t_{k-1}$, $\mathbb{U}_i(t_k)$ depends only on $\mathbb{U}_i(t_{k-1})$, etc., which implies $\mathbb{U}_i(t)$ is a continuous Markov process. The continuity follows the integral of equation (\ref{du=ndt}).

On the account that the random force $N_i(t)$ must average to
zero, if we choose $\mathbb{U}_i(0) = 0$ at the origin time, then $\langle \mathbb{U}_i(t_k) \rangle=0$ must hold.
Based on the discussion above, we may conclude that the increments of $\mathbb{U}_i$, 
\begin{equation}
    \mathbb{U}_i(t_1) - \mathbb{U}_i(t_0),\ \ \mathbb{U}_i(t_2) - \mathbb{U}_i(t_1),\ \ \cdots ,\mathbb{U}_i(t_k) - \mathbb{U}_i(t_{k-1}), \cdots \mathbb{U}_i(t_n) - \mathbb{U}_i(t_{n-1}) ~, \label{inc}
\end{equation}
are independent, stationary, and identically distributed with zero mean if the thermal motion in the medium has attained a steady state. 
By central limit theorem, we deduce that $\mathbb{U}_i(t)$ is a Gaussian process with zero mean. Thus, $\mathbb{U}_i(t)$ is a Wiener process scaled by $\sqrt{r}$, that is 
\begin{equation}
   \langle d \mathbb{U}_i(t) d\mathbb{U}_i(s) \rangle = r ( d t \cap d s )~. 
\end{equation}

By accounting for the particle thermal velocity in equilibrium, the magnitude of the random force $r$ can be written \cite{uhlenbeck1930theory} as
\begin{equation}
    r=2\gamma K_b T~,
\end{equation}
where $K_b$ is the Boltzmann constant and $T$ is the absolute temperature.
Then, equation (\ref{du=ndt}) can also be represented as
\begin{equation}
    d \mathbb{U}_i(t) = N_i(t) d t =  \sqrt{2\gamma K_b T} d \mathbb{W}_i(t) ~, \label{stoform}
\end{equation}
where $\mathbb{W}_i(t)$ is the standard Wiener process with zero mean and unit variance. This equation echos Doob's formula \cite{doob1942brownian} presenting the Langevin equation by a stochastic process. 
Also, the relation
\begin{equation}
    \langle d \mathbb{U}_i(t) d \mathbb{U}_j(t) \rangle  =  2\gamma K_b T \langle d \mathbb{W}_i(t) d \mathbb{W}_j(t)\rangle = 0
\end{equation}
holds since $d \mathbb{W}_i(t)$ is independent of $d \mathbb{W}_j(t)$ with $i$ and $j$ referring to the different coordinate indices of $x$, $y$ or $z$.

\section{The dynamics and diffusion of Brownian particles in general polynomial shear flows}   

We start the mathematical derivation by considering a two-dimensional laminar shear flow, of which the velocity profile may be described as a polynomial function of the transverse coordinate $y$, generally expressed here as
\begin{equation}
    v_f = \sum _{k=0}^{n} c_k U \left ( \frac{y}{L} \right )^k ~,  
\end{equation}
where $c_0, c_1, c_2, \cdots, c_n$ are dimensionless constant coefficients, $U$ the characteristic velocity of the flow, $L$ is the characteristic length scale of the flow, and $n$ is the order of flow velocity profile.
This representation will allow the generalization of the theory to resolve any shear flow that may be described by a polynomial function, either as a solution of the Navier-Stokes equations, such as Couette and Poisseuille flows, or as an approximation thereof.

To study the dynamics and diffusion of a Brownian particle carried by this polynomial laminar parallel flow, we may solve the Langevin equation (\ref{langevine}) in the stochastic form by defining 
\begin{equation}
    \textbf{U}_p =\left( \begin{array}{c} u_{p x} \\ u_{p y} \end{array} \right), \ \    \textbf{X}_p =\left( \begin{array}{c} x_{p} \\ y_{p} \end{array} \right),\ \ \textbf{V}_f = \binom{v_f}{0}, \ \
    \textbf{F}_{B}=\left( \begin{array}{c} 0 \\ 0 \end{array} \right),\ \  \textbf{N} =\left( \begin{array}{c} N_x \\ N_y \end{array} \right)~,
\end{equation}
where gravity and other body forces are neglected. 
The stochastic form of the two-dimensional polynomial laminar flow is then,
\begin{equation}
\left\{ \begin{array}{l} d \textbf{X}_p(t) = \textbf{U}_p(t) d t \\  \\ d \textbf{U}_p(t) =\frac{\gamma}{m} \Big ( \textbf{V}_f(t)-\textbf{U}_p(t)\Big ) d t + \frac{1}{m}\sqrt{2\gamma K_b T} d \mathbb{W}(t)\end{array} \right.  ~,        \label{differential_langevin}
\end{equation} 
where $\mathbb{W}(t)=\left( \begin{array}{c} \mathbb{W}_x(t) \\ \mathbb{W}_y(t) \end{array} \right)$.

Equation (\ref{differential_langevin}) is normalized by defining the following dimensionless variables:
\begin{equation}
    \Tilde{x}_p = \frac{x_p}{L},~~\Tilde{y}_p = \frac{y_p}{L},~~\Tilde{u}_{p x} = \frac{u_{p y}}{U},~~\Tilde{u}_{p y} = \frac{u_{p x}}{U},~~\Tilde{t} = \frac{t}{L/U},~~\Tilde{v}_{f} = \frac{v_f}{U} = \sum _{k=0}^{n} c_k \left ( \Tilde{y} \right )^k.  \label{nondim-varibles}
\end{equation}
We further define the Stokes number as $St = \frac{m/\gamma}{L/U}$, the ratio of particle relaxation time with flow characteristic time scale, and the dimensionless diffusion coefficient as $D^* = \frac{K_b T}{\gamma L U}$.  Note that $\sqrt{\frac{L}{U}}\mathbb{W}_i \left (\frac{t}{L/U} \right )$ is a standard Wiener process since $\mathbb{W}_i(t)$ is a standard Wiener process. Hence, $\sqrt{\frac{L}{U}}\mathbb{W}_i(\Tilde{t})$ is also 
 a standard Wiener process. Thus, 
\begin{equation}
  \left \langle d \mathbb{W}_i (\Tilde{t}) d \mathbb{W}_i (\Tilde{t}) \right \rangle = \frac{U}{L} d t = d \Tilde{t}, ~~ \left \langle d \mathbb{W}_i (\Tilde{t}) d \mathbb{W}_j (\Tilde{t}) \right \rangle = 0.  
\end{equation} 
Substituting the dimensionless variables in (\ref{nondim-varibles}) and henceforth omitting the tilde notations for convenience, we obtain the dimensionless governing equations,
\begin{equation}
\left\{ \begin{array}{l} d \textbf{X}_p(t) = \textbf{U}_p(t) d t ~; \\  \\ d \textbf{U}_p(t) =\frac{1}{St} \Big ( \textbf{V}_f(t)-\textbf{U}_p(t)\Big ) d t + \frac{1}{St}\sqrt{2D^*} d \mathbb{W}(t)~. \end{array}\right.  \label{Nond_stocha_Langevin} 
\end{equation} 


The analysis is divided into transverse and streamwise directions. The transverse velocity of the shear flow is assumed to be zero so that the particle dynamics and diffusion in the transverse direction are essentially the same as in quiescent flows. 
Under this assumption, by integrating the Langevin equation {(\ref{Nond_stocha_Langevin})} in the transverse direction, the particle position in the transverse direction is obtained as
\begin{equation}
y_p(t)= y_p(0) + St u_{py}(0)(1-e^{-\frac{t}{St}}) +  \sqrt{2 D^*}\int_{0}^t \Big( 1 - e^{-\frac{t-s}{St}} \Big )	~d\mathbb{W}_y(s)~.   \label{y_p}
\end{equation} 

In any polynomial shear flow, the particle's initial position $y_p(0)$ may change the particle diffusion depending on the local velocity and its gradient. If a particle is initially located at a higher (lower) flow velocity, the MSD will be larger (smaller) due to the higher (lower) flow velocity. However, only the gradient alters the variance of particle displacement, whereas the velocity affects the mean displacement. 
The initial velocity $u_{py}(0)$ is constant depending on the initial condition. It could be viewed either as the velocity at which the particle is introduced into the flow or the instantaneous velocity due to the Brownian motion. For a particle in thermal equilibrium, the Brownian motion in the transverse direction is the same as in quiescent flow. The particle speed is random and satisfies the 1-D Maxwell-Boltzmann distribution. Thus, the particle's initial velocity could be predicted by the root mean square velocity. By the energy equipartition theorem, the magnitude of $u_{py}(0)$ can be obtained as $\sqrt{\frac{k_B T}{m_p}}/U$ \cite{li2010measurement}. 
In the present study, the dimensionless variables, $y_p(0), u_{px}(0), u_{py}(0)$ and $c_k$ are considered of order $\mathcal{O}(1)$. The Stokes number, defined as the ratio of particle time scale to flow time scale, is assumed to be much less than $1$ for submicron particles within the shear flow.

Since the integral of the deterministic function $(1 -e^{-\frac{t-s}{St}})$ with respect to the standard Wiener process is a Gaussian process, the stochastic part of the particle position $\sqrt{2 D^*}\int_{0}^t \big( 1 - e^{-\frac{t-s}{St}}\big ) ~d \mathbb{W}_y(s)$ is a Gaussian process with zero mean and variance $\sigma^2_{y_p}$ in the following formula
\begin{equation}
     \sigma^2_{y_p} = 2 D^* \left[ \frac{St}{2}( 1 - e^{-\frac{2 t}{St}} ) + t - 2 St ( 1 - e^{-\frac{t}{St}} ) \right] ~.
\end{equation}
Given that $St$ is relatively small for Brownian particles, the variance of the particle position is asymptotic to $2 D^* t$ for time scales much longer than the $St$.
To maintain all the properties of the stochastic part of the particle position, as well as to simplify the problem, the stochastic term $- \sqrt{2 D^*} \int_{0}^t \big( 1 - e^{-\frac{t-s}{St}} \big ) ~d\mathbb{W}_y(s)$ may be approximated as $\sqrt{2D^*}\mathbb{W}_y(t)$ for time scales much longer than $St$.
Thus, for long-time scales, the particle position in the transverse direction can be written as
\begin{equation}
     y_p(t) \sim y_p(0) + St u_{py}(0)(1-e^{-\frac{t}{St}}) + \sqrt{2D^*}\mathbb{W}_y(t)  \label{y_p_asy}  ~.
\end{equation}


To solve equation (\ref{differential_langevin}) in the streamwise direction, we substitute $y_p(t)$ of equation (\ref{y_p_asy}) into the dimensionless flow velocity equation and obtain 
\begin{equation}
\begin{aligned}
    v_f(t) &= \sum _{k=0}^{n} c_k \left ( y_p(0) + St u_{py}(0)(1-e^{-\frac{t}{St}}) +  \sqrt{2 D^*} \mathbb{W}_y(t) \right)^k  \\&=
    \sum_{k=0}^{n} \sum_{\substack{\alpha, \beta, \lambda, \\ \alpha+\beta+\lambda=k}} c_k \binom{k}{\alpha, \beta, \lambda}  y_p^{\alpha}(0) St^{\beta} u_{py}^{\beta}(0) \left(1-e^{-\frac{t}{St}}\right)^{\beta} \left(\sqrt{2 D^*}\right)^{\lambda}\mathbb{W}_y^{\lambda}(t)        
\end{aligned}
\end{equation} 
with trinomial coefficients
\begin{equation*}
\binom{k}{\alpha, \beta, \lambda} =  \frac{k!}{\alpha!\beta!\lambda!}, ~~~~~~~~\alpha, \beta, \lambda \text{ are nonnegative integers}.
\end{equation*}
Denote $ \mathcal{F}(k,\alpha, \beta, \lambda) =  c_k \binom{k}{\alpha, \beta, \lambda}  y_p^{\alpha}(0) St^{\beta} u_{py}^{\beta}(0)\left(\sqrt{2 D^*}\right)^{\lambda}$, then $v_f(t)$ may be written as
\begin{equation}
    v_f(t) =
    \sum_{k=0}^{n} \sum_{\substack{\alpha, \beta, \lambda, \\ \alpha+\beta+\lambda=k}} \mathcal{F}(k,\alpha, \beta, \lambda) \left(1-e^{-\frac{t}{St}}\right)^{\beta} \mathbb{W}_y^{\lambda}(t) ~.           \label{FlowcV_exact_yp}
\end{equation}

By substituting (\ref{FlowcV_exact_yp}) into (\ref{Nond_stocha_Langevin}) and solving (\ref{Nond_stocha_Langevin}) in the streamwise direction, we obtain an expression for the particle velocity and displacement in the streamwise direction, 
\begin{equation}
    \begin{aligned}
      u_{px} &= 
      \sum_{k=0}^{n} \sum_{\substack{\alpha, \beta, \lambda, \\ \alpha+\beta+\lambda=k}} \mathcal{F}(k,\alpha, \beta, \lambda) St^{-1} \int_0^t e^{-\frac{t-t'}{St}}\left(1-e^{-\frac{t'}{St}}\right)^{\beta} \mathbb{W}_y^{\lambda}(t')  d t' + u_{px}(0) e^{-\frac{t}{St}}  + \frac{1}{St} \sqrt{2D^*}\int_0^t e^{-\frac{t-s}{St}} d \mathbb{W}_x(s)  
      \end{aligned}
\end{equation}
\begin{equation}
    \begin{aligned}
      s_{px} &= \sum_{k=0}^{n} \sum_{\substack{\alpha, \beta, \lambda, \\ \alpha+\beta+\lambda=k}} \mathcal{F}(k,\alpha, \beta, \lambda) \int_0^t \left(1- e^{-\frac{t-t'}{St}} \right)\left(1-e^{-\frac{t'}{St}}\right)^{\beta} \mathbb{W}_y^{\lambda}(t')  d t'  + St u_{px}(0) \left( 1-e^{-\frac{t}{St}} \right) \\& + \sqrt{2D^*}\int_0^t \left( 1- e^{-\frac{t-s}{St}} \right) d \mathbb{W}_x(s).  \label{exact_s_px}
    \end{aligned}
\end{equation}

To resolve the particle dynamics and diffusion in the streamwise direction, we first solve the Langevin equation (\ref{Nond_stocha_Langevin}) for the general polynomial laminar flow, analyze the particle diffusion in general polynomial laminar flows at long-time scales, and find the particle diffusion at all time scales for specific polynomial laminar flows.

\subsection{Long-time scale asymptotics} \label{HighOrderLF}

For long-time scales, $t\gg St$, all the exponential terms in equation (\ref{exact_s_px}) decay, and the particle displacement in the streamwise direction is asymptotic to:
\begin{equation}
       s_{px} \sim \sum_{k=0}^{n} \sum_{\substack{\alpha, \beta, \lambda, \\ \alpha+\beta+\lambda=k}} \mathcal{F}(k,\alpha, \beta, \lambda) \int_0^t \mathbb{W}_y^{\lambda}(t') d t'  + St u_{px}(0) + \sqrt{2D^*}\mathbb{W}_x(t).       
\end{equation}

Since the standard Wiener process is a normal distribution, by the moment generating function of the normal distribution and Taylor series, we have
\begin{equation}
E\big[ \mathbb{W}^{\lambda}(t)\big] = \left\{ \begin{array}{ccl}
0 & \mbox{for}
& \lambda=1,3,5,\dots(odd) \\ 
(\lambda-1)!! ~t^{\frac{\lambda}{2}} & \mbox{for} & \lambda=0,2,4,\dots(even)
\end{array}\right. ~.             \label{mean_power_WP}
\end{equation}

Then, the particle mean displacement at long-time scales is 
\begin{equation}
       \left \langle s_p(t) \right \rangle  \sim  \sum_{k=0}^{n}  \sum_{\substack{\alpha, \beta, \lambda, \\ \alpha+\beta+\lambda=k, \\ \lambda \text{ is even}}} \mathcal{F}(k,\alpha, \beta, \lambda) ( \lambda-1 )!!\frac{2}{\lambda +2} t^{\frac{\lambda+2}{2}}  + St u_{px}(0).  \label{mean-nth-long}
\end{equation}
The term $t^{\frac{\lambda+2}{2}}$ in equation (\ref{mean-nth-long}) arises from the expected value of the term with $\int_0^t  \mathbb{W}_y^{\lambda} (s) d s$ when $\lambda$ is even. That implies the variation of the particle displacement is offset by the integral of the odd powers of the Wiener process and is accumulated by the integral of the even powers of Wiener process.

Furthermore, by stochastic Fubini theorem and It$\hat{o}$'s lemma, the time integral of powers of Brownian motion can be represented as (see Appendix)
\begin{small}
\begin{equation}
\int_0^t  \mathbb{W}_y^{\lambda} (u) d u = \left\{ \begin{array}{lcl}
 \sum_{l=1}^{ \frac{\lambda+1}{2}} \frac{\lambda!}{2^{l-1}(\lambda-2l+1)!}\int_0^t  \frac{(t-s)^l}{l!}\mathbb{W}_y^{\lambda-(2l-1)}(s) d \mathbb{W}_y(s) & \mbox{for}
& \lambda=1,3,5,\dots(odd) \\ 
\\
\sum_{l=1}^{ \frac{\lambda}{2}} \frac{\lambda!}{2^{l-1}(\lambda-2l+1)!}\int_0^t  \frac{(t-s)^l}{l!}\mathbb{W}_y^{\lambda-(2l-1)}(s) d \mathbb{W}_y(s)  + (\lambda-1)!!\frac{2}{\lambda+2}t^{\frac{\lambda+2}{2}}& \mbox{for} & \lambda=2,4,6,\dots(even)
\end{array}\right. ~,            
\end{equation}
and for $\lambda = 0$, it holds that 
\begin{equation*}
    \int_0^t  \mathbb{W}_y^{\lambda} (u) d u = (\lambda-1)!!\frac{2}{\lambda+2}t^{\frac{\lambda+2}{2}} = t ~.
\end{equation*} 
\end{small}

In the case of $n =0$, the variance of the particle displacement is
\begin{equation}
    \sigma_{s_p}^2 = \left \langle \left ( \sqrt{2D^*}\mathbb{W}_x(t) \right)^2  \right \rangle = 2 D^* t ~, 
\end{equation}
which corresponds to particle diffusion in quiescent mediums. 
In the case of $n \geq 1$, the variance of the particle displacement for long-time scales is
\begin{small}
\begin{equation}
   \begin{aligned}
     \sigma_{s_p}^2  &= \left \langle \left ( \sum_{k=1}^{n} \sum_{\substack{\alpha, \beta, \lambda, \\ \alpha+\beta+\lambda=k, ~~\lambda > 0}} \mathcal{F}(k,\alpha, \beta, \lambda) \sum_{\zeta=1}^{\lceil \frac{\lambda}{2} \rceil} \frac{\lambda!}{2^{\zeta-1}(\lambda-2\zeta+1)!} \int_0^t \frac{(t-s)^\zeta}{\zeta!} \mathbb{W}_y^{\lambda-2\zeta+1}(s) d \mathbb{W}_y(s) +  \sqrt{2D^*}\mathbb{W}_x(t) \right)^2  \right \rangle    \\& =
     \sum_{k_1=1}^{n} \sum_{k_2=1}^{n} \sum_{\substack{\alpha_1, \beta_1, \lambda_1, \\ \alpha_1+\beta_1+\lambda_1=k_1, \\ \lambda_1 >0 }} \sum_{\substack{\alpha_2, \beta_2, \lambda_2, \\ \alpha_2+\beta_2+\lambda_2=k_2, \\ \lambda_2 >0 }} \mathcal{F}(k_1,\alpha_1, \beta_1, \lambda_1) \mathcal{F}(k_2,\alpha_2, \beta_2, \lambda_2) \sum_{\zeta_1=1}^{\lceil \frac{\lambda_1}{2} \rceil} \sum_{\zeta_2=1}^{\lceil \frac{\lambda_2}{2} \rceil} \left \langle  Z(\lambda_1,\zeta_1) Z(\lambda_2,\zeta_2) \right \rangle + 2 D^* t
     ~~~~\label{var-nth-long}       
    \end{aligned}
\end{equation}
with $\lceil \frac{\lambda}{2} \rceil$ a ceiling function giving the least integer greater than or equal to $\frac{\lambda}{2}$ and
\begin{equation}
   Z(\lambda,\zeta) = \frac{\lambda!}{2^{\zeta-1}(\lambda-2\zeta+1)!} \int_0^t \frac{(t-s)^\zeta}{\zeta!} \mathbb{W}_y^{\lambda-2\zeta+1}(s) d \mathbb{W}_y(s) ~.
\end{equation}
\end{small}

By It$\hat{o}$ isometry, $\left \langle  Z(\lambda_1,\zeta_1) Z(\lambda_2,\zeta_2) \right \rangle$ may be presented as
\begin{equation}
\left \langle  Z(\lambda_1,\zeta_1) Z(\lambda_2,\zeta_2) \right \rangle =\mathcal{G} \int_0^t  \frac{(t-s)^{\zeta_1+\zeta_2}}{\zeta_1!~\zeta_2!} \big \langle \mathbb{W}_y^{\mathcal{J}}(s) \big \rangle d s
\end{equation}
with $\mathcal{J} = \lambda_1+\lambda_2-2(\zeta_1+\zeta_2)+2 $ and $$\mathcal{G} = \frac{ \lambda_1!~\lambda_2!}{2^{\zeta_1+\zeta_2-2}(\lambda_1-2\zeta_1+1)!(\lambda_2-2\zeta_2+1)!}~.$$

Then, by the properties of the Wiener process and equation (\ref{mean_power_WP}), we obtain
 \begin{equation}
\big \langle Z(\lambda_1,\zeta_1) Z(\lambda_2,\zeta_2) \big \rangle = \left\{ \begin{array}{cl}
0  & ~~~~~\mathcal{J} ~ is ~ odd \\
\\
    \mathcal{G} \sum_{i=0}^{\zeta_1+\zeta_2} \binom{\zeta_1+\zeta_2}{i} (-1)^{\zeta_1+\zeta_2-i} \frac{2}{\lambda_1+\lambda_2+4-2i}   t^{\frac{\lambda_1+\lambda_2+4}{2}} & ~~~~~\mathcal{J} ~ is ~ even  
\end{array}\right. ~,
\end{equation}
with binomial coefficient $\binom{\zeta}{i} = \frac{(\zeta)!}{i!(\zeta-i)!}$.
Moreover, $\big \langle Z(\lambda_1,\zeta_1) Z(\lambda_2,\zeta_2) \big \rangle$ receives its highest power, denoted by $\big \langle Z(\lambda_1,\zeta_1) Z(\lambda_2,\zeta_2) \big \rangle_h$, when both $\lambda_1$ and $\lambda_2$ are $n$. That is, 
\begin{equation}
    \big \langle Z(\lambda_1,\zeta_1) Z(\lambda_2,\zeta_2) \big \rangle_h =\mathcal{G} \sum_{i=0}^{\zeta_1+\zeta_2} \binom{\zeta_1+\zeta_2}{i} (-1)^{\zeta_1+\zeta_2-i} \frac{1}{n+2-i} t^{n+2} ~.
\end{equation} 

The variance of the particle displacement in equation (\ref{var-nth-long}) is a polynomial function of $t$. The term with highest power, $t^{n+2}$, in equation (\ref{var-nth-long}) emerges from the variance of the term $\int_0^t  \mathbb{W}_y^n (s) d s$.
The second highest power, $t^{n+1}$, results from the variance of the term $\int_0^t \mathbb{W}_y^{n-1} (s) d s$ or the covariance between terms ~~$\int_0^t\mathbb{W}_y^n (s) d s$~~and~~$\int_0^t\mathbb{W}_y^{n-2} (s) d s$. 
So do the other powers of $t$. It implies that the combination of the flow velocity function, shown in the exponents, and the Brownian motion in the transverse direction $\mathbb{W}_y$, shown in the base, makes a remarkable difference to the variance of the particle displacement.

The particle MSD at long-time scales can be calculated by the following formula, $\langle s_p^2 \rangle = \langle s_p \rangle^2 + \sigma_{s_p}^2$.
For odd $n$, the highest power of $t$ in $\langle s_p \rangle^2$ is $n+1$ since only the integral of even power of the Wiener process accumulates in the mean displacement, while the highest power of $t$ in $\sigma_{s_p}^2$ is $n+2$. For even $n$, the highest power of $t$ in both $\langle s_p \rangle^2$ and $\sigma_{s_p}^2$ are $n+2$.
Thus, we obtain, at long-time scales, particle MSD in the streamwise direction is a polynomial function of $t$ with the highest power $n+2$. In other words, at long-time scales, the order of the particle diffusion is $n+2$ for the order of flow velocity $n$. 

The summation term in equation (\ref{var-nth-long}) characterizes the dominant term of particle diffusion in the streamwise direction due to the coupling of particle diffusion and flow velocity gradient in the transverse direction. The second term, $2 D^* t$, reveals particle pure diffusion in the streamwise direction.


To verify the formulas for the general velocity profile, we compare our present results with the results from Foister et al. \cite{foister1980diffusion} for the cases of Couette and plane-Poiseuille flows. In their study, instead of using a space-fixed coordinate system, the authors defined a new coordinate system in which the flow velocity at the particle's initial position $v_f(0)$ is subtracted. Thus, for the comparison, we subtract a displacement due to $v_f(0)$, in order to verify the present MSD.

For Couette flow, $n = 1$, $c_0=c_2=\cdots=c_n=0$, and $v_f(0) = c_1 y_p(0)$. Subtracting $c_1 y_p(0) t$ from $s_p$ and omitting the terms with all powers of $St$ for the limit $t \to \infty$, we obtain the asymptotic MSD as
\begin{equation}
    \big \langle s_p^2 \big \rangle  =  2 D^* t \left( 1 + c_1^2 \frac{t^2}{3}\right) ~,
\end{equation}
which is the same as the results reported in  \cite{foister1980diffusion}.

For plane-Poiseuille flow, $n=2$, $c_0 = 1$, $c_1 = 0,~ c_2 = - 1$, $c_3=c_4=\cdots=c_n=0$, $v_f(0) = c_0 + c_2y_p^2(0) $. Subtracting $\left( c_0 + c_2y_p^2(0) \right) t$ from $s_p$ and omitting the terms with all powers of $St$ for the limit $t \to \infty$, we obtain the asymptotic MSD for the plane-Poiseuille flow as
\begin{equation}
    \big \langle s_p^2 \big \rangle  = 2 D^* t \left( 1 + c_2^2 y_p^2(0) \frac{4 t^2}{3} + D^* c_2^2 \frac{7 t^3}{6} \right)  ~,
\end{equation}
which is the exact same result reported by Foister et al. \cite{foister1980diffusion}.

\begin{figure}[htbp]
\centering
\includegraphics[width=16cm]{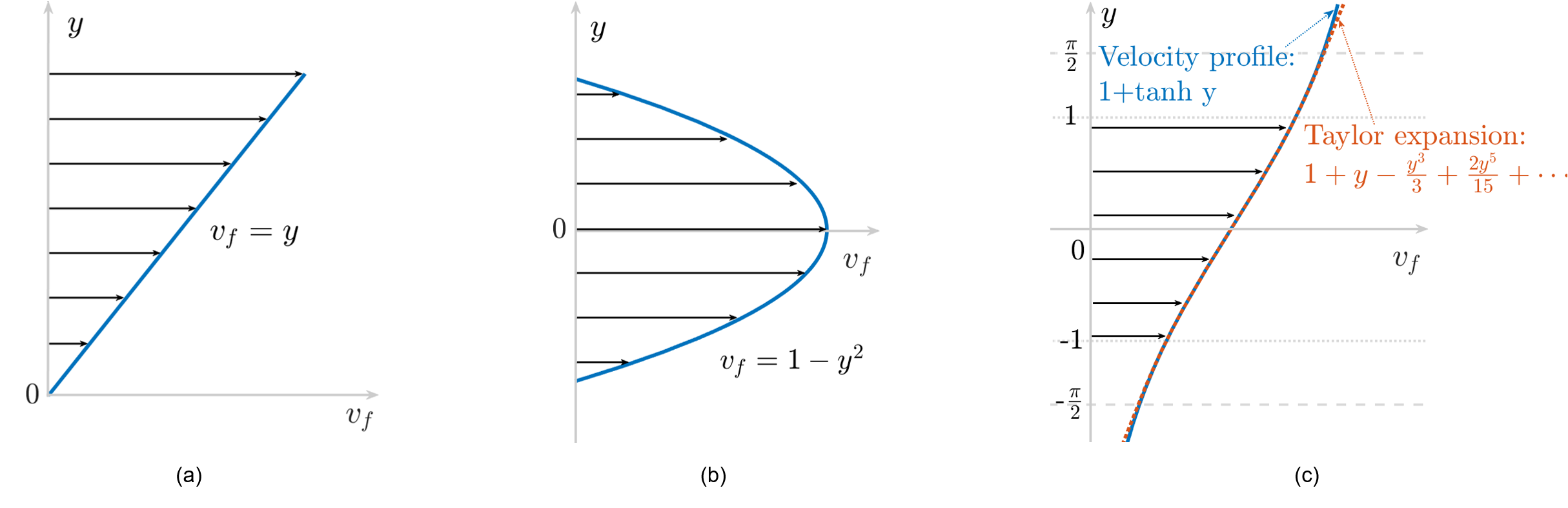}
\caption{Velocity profile of (a) Couette flow, (b) plane-Poiseuille flow, (c) hyperbolic tangent flow with the Taylor expansion at origin presented in equations (\ref{Couettefun}), (\ref{planepfun}), and (\ref{htpfun}) respectively.}
\label{fig-linear-parab}
\end{figure}

In the following sections, using the generalized theory developed here, we shall explore the diffusion of particles in three specific test cases, illustrated here in figure~\ref{fig-linear-parab}. 

\subsection{Brownian particle diffusion in Couette flows at all time regions}  \label{CouetteF}

We begin the analysis by expanding the calculation of particle diffusion in Couette flows for all time regions. Specifically, the particle diffusion will be resolved for the short, long, and intermediate-time scales.
A dimensionless Couette flow, illustrated in Figure \ref{fig-linear-parab} (a), and described by
\begin{equation}
 v_f(t)  =  y  \label{Couettefun}
\end{equation}
is considered here, where the effects of boundaries on particle diffusion are neglected.

Here, $c_0=c_2=c_3=\cdots=c_n=0$; $c_1 =1$ and the particle initial position is $ \big( x_p(0), y_p(0) \big )$. 
The displacement in equation (\ref{exact_s_px}) for this unbounded Couette flow is
\begin{equation}
    \begin{aligned}
      s_{px} &= \sum_{k=0}^{1} \sum_{\substack{\alpha, \beta, \lambda, \\ \alpha+\beta+\lambda=k}} \mathcal{F}(k,\alpha, \beta, \lambda) \int_0^t \left(1- e^{-\frac{t-s}{St}} \right)\left(1-e^{-\frac{s}{St}}\right)^{\beta} ~\mathbb{W}_y^{\lambda}(s) d s  + St u_{px}(0) \left( 1-e^{-\frac{t}{St}} \right) \\& + \sqrt{2D^*}\int_0^t \left( 1- e^{-\frac{t-s}{St}} \right) d \mathbb{W}_x(s). 
    \end{aligned}
\end{equation}

Using It$\hat{o}$'s lemma, we may write
$ \int_0^t  \mathbb{W}_y (u) d u   = \int_0^t  (t-s) d\mathbb{W}_y(s)$ (see Appendix).
Then, the particle displacement takes the form,
\begin{equation}
    \begin{aligned}
      s_{px} &= c_1 y_p(0) \left[ t - St \left( 1 - e^{-\frac{t}{St}}\right)\right] + c_1 St u_{py}(0) \left[ \left( 1 + e^{-\frac{t}{St}}\right) t - 2 St \left( 1 - e^{-\frac{t}{St}}\right)\right] + St u_{px}(0) \left( 1-e^{-\frac{t}{St}} \right)  \\
      &+ c_1 \sqrt{2 D^*} \left[ \int_0^t \left( (t - s) + St \left( e^{-\frac{t - s}{St}} - 1\right)\right) d~ \mathbb{W}_y(s) \right ]+ \sqrt{2D^*}\int_0^t \left( 1- e^{-\frac{t-s}{St}} \right) d \mathbb{W}_x(s)   ~.  
    \end{aligned}
\end{equation}
Finally, by taking the expected value of the particle displacement and calculating the variance, we have 

\begin{small}
\begin{equation}
\begin{aligned}
    \big \langle s_p \big \rangle &=  c_1 y_p(0) \left[ t - St \left( 1 - e^{-\frac{t}{St}}\right)\right] + c_1 St u_{py}(0) \left[ \left( 1 + e^{-\frac{t}{St}}\right) t - 2 St \left( 1 - e^{-\frac{t}{St}}\right)\right] + St u_{px}(0) \left( 1-e^{-\frac{t}{St}} \right)  ~;
    \\
    \sigma_{s_p}^2  & = 2 D^* \left \{ c_1^2 \left [ \frac{1}{3}t^3  - t^2 St  - t St^2 + 2 t St^2 \left( 1 - e^{-\frac{t}{St}}\right) + \frac{1}{2} St^3 \left( 1 - e^{-\frac{2t}{St}}\right)\right] \right\} + 2 D^* \left[ \frac{St}{2} (1-e^{-\frac{2t}{St}}) -2 St (1-e^{-\frac{t}{St}}) +t \right]~.     \label{mean-var-linear}
\end{aligned}
\end{equation}
The particle MSD is obtained by calculating $\left \langle s_p^2 \right \rangle = \left \langle s_p \right \rangle^2 + \sigma_{s_p}^2$. 
Note that the formula converges to the particle MSD in quiescent flow when the shear rate $c_1$ is zero. 
\end{small}

\subsubsection{Short-time scale analysis}  \label{linear-short}

For time scales much shorter than the Stokes number $St$, that is $t \ll St \ll 1 $, $\frac{t}{St} \ll 1$ holds. Expanding $e^{-\frac{t}{St}}$ and $e^{-\frac{2t}{St}}$ with third-order Taylor polynomial series around $0$, we approximate the particle MSD as
\begin{equation}
    \big \langle s_p^2 \big \rangle \sim   u_{p x}^2(0) t^2  ~.\label{msd-linear-short}
\end{equation} 
If the initial velocity of the Brownian particle in the streamwise direction is only the velocity due to the random motion. Taking a second average over $u_{p x}(0)$, by thermal equilibrium, we have
\begin{equation}
   \big \langle \big \langle s_p^2 \big \rangle \big \rangle \sim   \big \langle u_{p x}^2(0) \big \rangle t^2 = \frac{K_b T}{m_p} t^2  ~.     \label{short-ThermalEquli}
\end{equation}
It shows that, on average, the particle has a uniform motion at short-time scales and that the MSD is the same as it is in quiescent flow or stationary medium \cite{uhlenbeck1930theory} since time is too short for the particle to catch up with the flow velocity. Moreover, the transverse coordinate of the particle changes so little during a very short period of time that the flow velocity does not vary enough to make any difference to the particle diffusion in the streamwise direction.

\subsubsection{Long time-scale analysis}  \label{linear-long}   

When the normalized time $t$ is much longer than the Stokes number $St$, 
$e^{-\frac{t}{St}} \to 0$, and $e^{-\frac{2t}{St}} \to 0$ hold; the leading terms of the particle mean displacement, the variance of particle displacement, and the particle MSD can be approximated by
\begin{equation}
\begin{aligned}
    \big \langle s_p \big \rangle &\sim   c_1 y_p(0)  t  ;
   ~~~~~~~
    \sigma_{s_p}^2 \sim 2 D^* t\big ( c_1^2 \frac{t^2}{3} +1  \big )  ;
       \\
    \big \langle s_p^2 \big \rangle &\sim  \left[ c_1 y_p(0) t \right]^2 +  2 D^* t\big ( c_1^2 \frac{t^2}{3} +1  \big )    ~. \label{msd-linear-long} 
\end{aligned}
\end{equation}
On average, the particle travels at a velocity equal to the flow velocity at the particle's initial position $c_1 y_p(0)$. The reasons may be illustrated as follows. On the one hand, there is enough time for the particle to catch up with the flow velocity. On the other hand, the Brownian motion in the transverse direction results in particle hopping between different velocity layers, while the linearity of the flow velocity offsets the variation of the particle mean displacement in the streamwise direction.
Moreover, the variance of the particle displacement in the streamwise direction is larger than that in quiescent flows by $2 D^* c_1^2 \frac{t^3}{3}$, which comes from the variance of the term $\int_0^t \mathbb{W}_y (s) ds$. This enhanced diffusion is the same as the results obtained in \cite{foister1980diffusion, katayama1996brownian}. This allows us to mathematically perceive the effects of the combination of Brownian motion in the transverse direction $\mathbb{W}_y$ and the linear velocity function, shown in the power of $\mathbb{W}_y$, on the particle diffusion in the streamwise direction.

Hence, for this linear case, the coupling between the Brownian motion in the transverse direction and the flow velocity does not alter the mean displacement but rather the variance of particle displacement in the streamwise direction.

\begin{figure}[htbp]
\centering
\begin{minipage}[t]{0.3\textwidth}
\centering
\includegraphics[width=5.3cm]{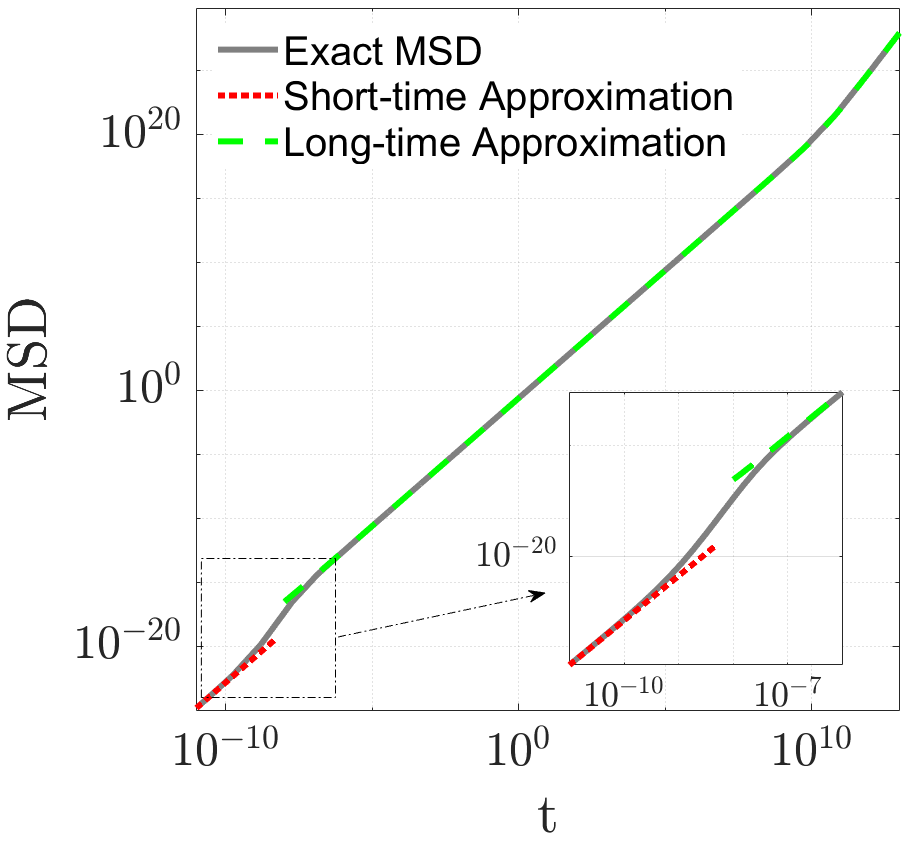}
\centerline{(a)}
\end{minipage}
\hspace{0.2in}
\begin{minipage}[t]{0.3\textwidth}
\centering
\includegraphics[width=5.3cm]{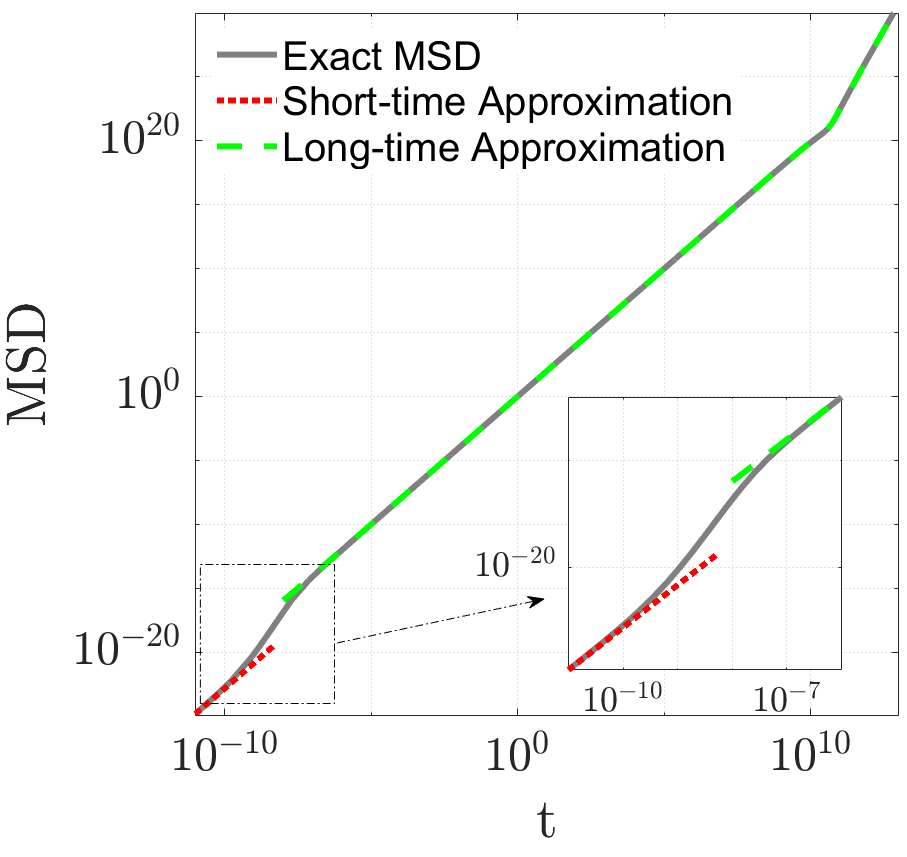}
\centerline{(b)}
\end{minipage}
\hspace{0.2in}
\begin{minipage}[t]{0.3\textwidth}
\centering
\includegraphics[width=5.3cm]{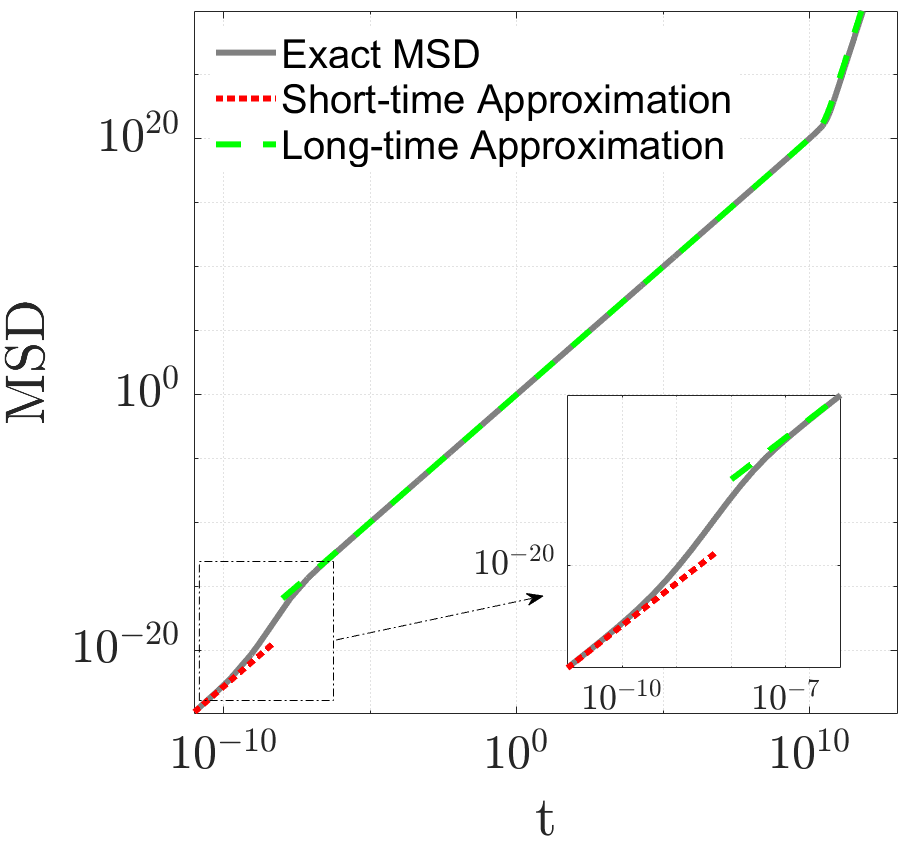}
\centerline{(c)}
\end{minipage}
\caption{Comparison of the exact MSD with short-time and long-time approximation (a) Couette flow, (b) plane-Poiseuille, and (c) hyperbolic tangent flow, respectively. The gray curves are the exact particle MSD in each flow, the green dashed curves are the corresponding long-time approximated MSD, and the red dotted curves are the corresponding short-time approximated MSD.}
\label{fig-comp-SL}
\end{figure}

We demonstrate the particle MSD for particles initially located at $y = 0.5$ with $St = 10^{-8}$ at $20^{\circ}C$ in Figure \ref{fig-comp-SL}(a). It shows that the particle approximated MSD at long-time scales in equation (\ref{msd-linear-long}) and the particle approximated MSD at short-time scales in equation (\ref{msd-linear-short}) is in good agreement with the exact MSD obtained from equation (\ref{mean-var-linear}). 

It should be noted that the particle MSD at short-time scales is known (see section (\ref{linear-short})), and particle diffusion at long-time scales has been discussed in the literature \cite{foister1980diffusion, katayama1996brownian}. However, our result continuously bridges these two approximations and reveals the exact particle MSD in Couette flows for all time scales.
 
\subsection{Brownian particle diffusion in plane-Poiseuille flows at all time regions}  \label{PPoiseF}

We shall now consider a 2D plane-Poiseuille flow with a dimensionless parabolic velocity profile (see Figure \ref{fig-linear-parab} (b)), described by the equation
\begin{equation}
    v_f =  1 - y^2  ~.  \label{planepfun}
\end{equation}
Here, as in the linear Couette flow case, we neglect the effect of boundaries on the diffusion of particles. 
$y$ is the dimensionless vertical coordinate with the origin at the centerline. Then, $c_0 = 1, ~c_1 = 0,~ c_2 = - 1, ~c_3=c_4=\cdots=c_n=0$. 
The displacement in equation (\ref{exact_s_px}) for this unbounded plane-Poiseuille flow is
\begin{equation}
    \begin{aligned}
      s_{px} &= \sum_{k=0}^{2} \sum_{\substack{\alpha, \beta, \lambda, \\ \alpha+\beta+\lambda=k}} \mathcal{F}(k,\alpha, \beta, \lambda) \int_0^t \left(1- e^{-\frac{t-s}{St}} \right)\left(1-e^{-\frac{s}{St}}\right)^{\beta} ~\mathbb{W}_y^{\lambda}(s) d s  + St u_{px}(0) \left( 1-e^{-\frac{t}{St}} \right) \\& + \sqrt{2D^*}\int_0^t \left( 1- e^{-\frac{t-s}{St}} \right) d \mathbb{W}_x(s). 
    \end{aligned}
\end{equation}

Using It$\hat{o}$'s lemma and It$\hat{o}$ isometry, we may write
$ \mathbb{W}_y^2 (t) = \int_0^t 2 \mathbb{W}_y(s)d\mathbb{W}_y (s)  +  t$
and $ \int_0^t  \mathbb{W}_y^2 (u) d u   = \int_0^t 2 (t-s)\mathbb{W}_y(s)  d\mathbb{W}_y(s) + \frac{t^2}{2}  $ (see Appendix).
Then, the particle displacement becomes

\begin{small}
\begin{equation}
\begin{aligned}
          s_p(t) &= \left( c_0 +c_2 y_p^2(0)\right)\left[ t - St \left( 1-e^{-\frac{t}{St}} \right) \right] + 2 c_2 y_p(0) St u_{py}(0) \left[ \left( 1 + e^{-\frac{t}{St}} \right) t - 2 St \left( 1-e^{-\frac{t}{St}} \right) \right] \\
          & + 2 c_2 y_p(0) \sqrt{2 D^*} \int_0^t \left[ (t - s) - St \left( 1 - e^{-\frac{t - s}{St}} \right) \right] ~d \mathbb{W}_y(s) + c_2 St^2 u_{py}^2(0) \left[ \frac{St}{2} e^{-\frac{2 t}{St}} + 2(t+ St) e^{-\frac{t}{St}} + \left( t- \frac{5}{2}St \right) \right] \\
          &+ 2 c_2 St u_{py}(0) \sqrt{2 D^*} \int_0^t \left[ \left( 1 - e^{-\frac{t}{St}} \right)(t - s) - St \left( 1 - e^{-\frac{t - s}{St}} \right) \left( 1 + e^{-\frac{s}{St}} \right) \right] ~d \mathbb{W}_y(s)\\
          &+ 2 c_2 D^* \int_0^t \left[ (t - s) - St \left( 1 - e^{-\frac{t - s}{St}} \right) \right]2\mathbb{W}_y(s) ~d \mathbb{W}_y(s) + 2 c_2 D^* \left[ \frac{t^2}{2} - tSt + St^2 \left( 1-e^{-\frac{t}{St}} \right) \right] \\
          & + St u_{px}(0) \left( 1-e^{-\frac{t}{St}} \right) + \sqrt{2 D^*} \int_0^t \left( 1 - e^{-\frac{t - s}{St}} \right)  ~d \mathbb{W}_x(s)~.
\end{aligned}
\end{equation}
\end{small}

Taking the expected value of the particle displacement and its variance, we have
\begin{small}
\begin{equation}
\begin{aligned}
\big \langle s_p(t) \big \rangle 
  &= \left( c_0 +c_2 y_p^2(0)\right)\left[ t - St \left( 1-e^{-\frac{t}{St}} \right) \right] + 2 c_2 y_p(0) St u_{py}(0) \left[ \left( 1 + e^{-\frac{t}{St}} \right) t - 2 St \left( 1-e^{-\frac{t}{St}} \right) \right]  + St u_{px}(0) \left( 1-e^{-\frac{t}{St}} \right) \\
  &+ c_2 St^2 u_{py}^2(0) \left[\left( t- \frac{5}{2}St \right) +  2(t+ St) e^{-\frac{t}{St}} + \frac{St}{2} e^{-\frac{2 t}{St}} \right] + 2 c_2 D^* \left\{ \frac{t^2}{2} - tSt + St^2 \left( 1-e^{-\frac{t}{St}} \right) \right\}~;   
\\
  \sigma_{s_p}^2 & = 8 c_2^2 y_p^2(0) D^* \left[\left(\frac{1}{3} t^3 - t^2 St + t St^2 +\frac{1}{2} St^3 \right) - 2 t St^2 e^{-\frac{t}{St}} - \frac{1}{2} St^3 e^{-\frac{2t}{St}}\right] \\
  &+8 c_2^2 St^2 u_{py}^2(0) D^* \left[\left(\frac{1}{3} t^3 - t^2 St - t St^2 + 5 St^3 \right) + \left( \frac{2}{3}t^3 - 8 t St^2 \right) e^{-\frac{t}{St}} + \left( \frac{1}{3}t^3 + t^2 St - t St^2 - 5St^3 \right) e^{-\frac{2t}{St}}\right] \\
  & + 16~c_2^2~(D^*)^2 \left[\left(\frac{1}{12} t^4 - \frac{1}{3}t^3 St + \frac{1}{2}t^2 St^2 + \frac{1}{2}t St^3 - \frac{9}{4} St^4 \right) + \left(2 t St^3 + 2 St^4 \right) e^{-\frac{t}{St}} + \frac{1}{4} St^4 e^{-\frac{2t}{St}}\right] \\
  &+16 c_2^2 y_p(0) St u_{py}(0) D^* \left[\left(\frac{1}{3} t^3 - t^2 St + \frac{5}{2} St^3 \right) + \left( \frac{1}{3}t^3 - 4t St^2 \right) e^{-\frac{t}{St}} - \left( t St^2 + \frac{5}{2}St^3 \right) e^{-\frac{2t}{St}}\right] \\
  &+ 2 D^* \left[ (t -\frac{3}{2} St) + 2St e^{-\frac{t}{St}} - \frac{1}{2} St e^{-\frac{2t}{St}} \right]~.   \label{mean-var-para}
\end{aligned}
\end{equation}
\end{small}
And the particle MSD for all time regions can now be obtained by calculating $\left \langle s_p^2 \right \rangle = \left \langle s_p \right \rangle^2 + \sigma_{s_p}^2$.

\subsubsection{Short time-scale limit}  \label{para-short}    

For time much shorter than the Stokes number $St$, that is $t \ll St \ll 1$, $\frac{t}{St} \ll 1$ holds. Taking the first-order Taylor polynomial approximation to $e^{-\frac{t}{St}}$ and $e^{-\frac{2t}{St}}$ around $0$, we approximate the particle MSD as
\begin{equation}
    \big \langle s_p^2 \big \rangle \sim   u_{p x}^2(0) t^2  ~,         \label{msd-para-short}
\end{equation}
which is the same expected expression obtained here for particle diffusion at short-time scales in Couette flow. When the particle's initial velocity is only the velocity due to the random motion, equation (\ref{short-ThermalEquli}) is also revealed in thermal equilibrium by taking a second average over $u_{p x}(0)$.

From the analysis of short-time scales for Couette flow in section \ref{linear-short} and plane-Poiseuille flow above, we deduce the expected result that no matter what the laminar flow velocity is, the particle travels with its initial velocity at short-time scales, and the particle diffusion is the same as in quiescent media. 
Nevertheless, this comparison is valuable for the validity of our stochastic approach in the limit of short-time scales. 

\subsubsection{Long time-scale limit} \label{para-long}    

For time much longer than the Stokes number $St$, 
$e^{-\frac{t}{St}}\to 0$, and $e^{-\frac{2t}{St}} \to 0$ hold. Then, the leading terms of the particle mean displacement, the variance of the particle displacement, and the MSD are  expressed by 
\begin{equation}
\begin{aligned}
    \big \langle s_p \big \rangle &\sim  \left( c_0 +c_2 y_p^2(0)\right) t + c_2 D^*  t^2; ~~~~~~~
   \sigma_{s_p}^2  \sim 2 D^* t \Big [ c_2^2 D^* \frac{2 t^3}{3} + c_2^2 y_p^2(0) \frac{4 t^2}{3} + 1  \Big ] ;
   \\
   \big \langle s_p^2 \big \rangle &\sim  \left[\left( c_0 +c_2 y_p^2(0)\right) t + c_2 D^*  t^2 \right]^2 
+ 2 D^* t \Big [ c_2^2 D^* \frac{2 t^3}{3} + c_2^2 y_p^2(0) \frac{4 t^2}{3} + 1  \Big ]   \label{msd-para-long} ~.
\end{aligned}
\end{equation}

For long-time scales, the particle travels at a larger velocity than the flow velocity at the particle's initial position $c_0 + c_2 y_p^2(0)$. 
Due to the coupling between random motion in the transverse direction and the parabolic velocity profile of the flow, the particle velocity increases incrementally compared to $c_0 + c_2 y_p^2(0)$. Here, the increase results in an extra term (that is $c_2 D^*  t^2$) in the formula of the mean displacement. If $c_2$ is negative, the particle velocity is lower than the flow velocity at the particle's initial position. 
Moreover, the highest power of $t$ of the variance in $4 {D^*}^2 c_2^2 t^4/3$ emerges from calculating the variance of the term $2 c_2 D^* \int_0^t\mathbb{W}_y^2(s) d s$. This also gives us a glimpse - through the mathematical formulation - of how the combination of random motion in the transverse direction $\mathbb{W}_y$ and the parabolic velocity function, shown in the power of $W_y(t)$, impact the variance of particle displacement in the streamwise direction.

Hence, for this second-order polynomial velocity case, the coupling between the random motion in the transverse direction and the flow velocity gradient drastically changes not only the variance of particle displacement in the streamwise direction but also the particle mean displacement. Suppose a group of particles is released freely and independently into the unbounded plane-Poiseuille flow at the same position; the mean displacement of all the particles will be $\left( c_0 +c_2 y_p^2(0)\right) t + c_2 D^*  t^2 $ instead of $\left( c_0 +c_2 y_p^2(0)\right) t$, and particles will disperse in the streamwise direction with a standard deviation $\sqrt{2 D^* t \Big [ c_2^2 D^* \frac{2 t^3}{3} + c_2^2 y_p^2(0) \frac{4 t^2}{3} + 1  \Big ]}$, which is greater than the classical diffusion $\sqrt{2 D^* t}$ by a factor of $\sqrt{ c_2^2 D^* \frac{2 t^3}{3} + c_2^2 y_p^2(0) \frac{4 t^2}{3} + 1 }$.

Here, we demonstrate the particle MSD for particles initially located at $y = 0$ with $St = 10^{-8}$ at $20^{\circ}C$ in Figure \ref{fig-comp-SL}(b). It shows that the particle approximated MSD at long-time scales in equation (\ref{msd-para-long}) and the particle approximated MSD at short-time scales in equation (\ref{msd-para-short}) fit well with the exact MSD obtained from equations in (\ref{mean-var-para}). 
Our results unify these two approximations and reveal the particle MSD in plane-Poiseuille flow for all time regions.

\subsection{Brownian particle diffusion in a hyperbolic tangent flow}  \label{PHTF}

In this section, we demonstrate how the stochastic generalized method developed here for Brownian particle diffusion may theoretically be applied to any two-dimensional shear flow under the limitation of polynomial approximation.
We can approximate the flow velocity by expanding its profile into a polynomial series and studying the spatial region around $y_p(0)$ for which the polynomial is accurate enough and fits the original flow velocity profile. 

To demonstrate this approach, we consider a two-dimensional shear flow, analytically described by a hyperbolic tangent flow with a velocity profile,
\begin{equation}
    v_f = 0.5 \left( 1 + \tanh \left( \frac{y}{\delta} \right)\right)  ~,
\end{equation}
where $\delta$ is the shear thickness. Let Brownian particles be released freely and independently into the flow at $y_p(0) = 0$. We may expand $\tanh \left( y/\delta \right)$ at $y=0$ and investigate the particle diffusion in the flow region between $y = \delta \pi /2$ and $y = -\delta \pi /2$.

We choose $\Tilde{v}_f = v_f/0.5$, $\Tilde{y} = y/\delta$ for normalization, and drop all the tilde notation for convenience, and obtain the dimensional velocity profile (see Figure \ref{fig-linear-parab}c) as
\begin{equation}
    v_f =  1 + \tanh(y)  ~. 
\end{equation}
Expanding $\tanh(y) $ by Taylor series around $y=0$, we obtain the flow velocity in a polynomial formula in the region between $y=\pi/2$ and $y=-\pi/2$ as
\begin{equation}
    v_f =  1 + y -\frac{y^3}{3} + \frac{2y^5}{15} + \cdots ~.  \label{htpfun}
\end{equation}
For convenience, we investigate the particle diffusion between $y=1$ and $y=-1$, shown in Figure \ref{fig-linear-parab}. 
The effects of these artificial boundaries and the order to which the polynomial function is expanded will be discussed in section \ref{disc}.

Here, we shall consider the first four terms for the expansion of $v_f$. Then, the coefficients $c_0 = 1, ~c_1 = 1,~ c_2 = 0, ~c_3 = -1/3;~c_4 = 0;~c_5 = 2/15;~c_6=c_7=\cdots=c_n=0$ are determined. Hence, the particle displacement in equation (\ref{exact_s_px}) for this hyperbolic tangent flow is
\begin{equation}
    \begin{aligned}
      s_{px} &= \sum_{k=0}^{5} \sum_{\substack{\alpha, \beta, \lambda, \\ \alpha+\beta+\lambda=k}} \mathcal{F}(k,\alpha, \beta, \lambda) \int_0^t \left(1- e^{-\frac{t-s}{St}} \right)\left(1-e^{-\frac{s}{St}}\right)^{\beta} ~\mathbb{W}_y^{\lambda}(s) d s  + St u_{px}(0) \left( 1-e^{-\frac{t}{St}} \right) \\& + \sqrt{2D^*}\int_0^t \left( 1- e^{-\frac{t-s}{St}} \right) d \mathbb{W}_x(s). 
    \end{aligned}
\end{equation}

Following the same steps taken in the cases of Couette and plane-Poiseuille flows, we calculate the mean and variance of the particle displacement and then the exact MSD. For concision, we shall not present the equations here due to the long formulas. Rather, we report the final results for the exact MSD, the expected short-time approximated MSD, and the long-time approximated MSD obtained from equations (\ref{mean-nth-long}) and (\ref{var-nth-long}) in Figure \ref{fig-comp-SL}(c). In this case, particles are also initially located at $y = 0$ with $St = 10^{-8}$ at $20^{\circ}C$. As expected, the deduction about particle MSD at short-time scales is also verified in Figure \ref{fig-comp-SL}(c). It shows the coincidence of the exact MSD and the leading terms of particle diffusion at long-time scales.

\begin{figure}[htbp]
\centering
\centering
\includegraphics[width=8cm]{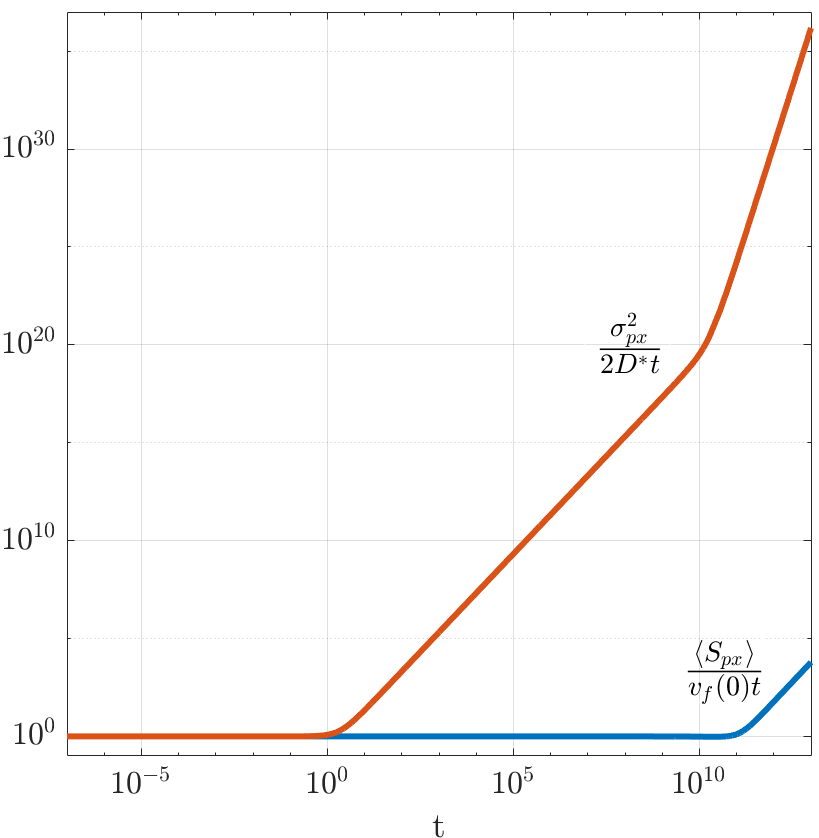}
\caption{Increase of the particle mean displacement and variance in Hyperbolic tangent flow.}
\label{InHT}

\end{figure}

We denote the flow velocity at the initial position of the particle as $v_f(0)$. In this case, $v_f(0) = 1 + y_p(0) -\frac{y_p^3(0)}{3} + \frac{2y_p^5(0)}{15}$. As time proceeds much further than $St$, the increase of particle mean displacement compared to $v_f(0) t$ and the increase of the variance of particle displacement compared to $2D^*t$ can also be estimated, shown in Figure (\ref{InHT}).

Figure \ref{InHT} conveys that the particle mean displacement remains in the order of $v_f(0)t$ for a relatively long time and grows as time proceeds. However, the variance of particle displacement stays in the order of $2D^*t$ at first. Thereafter, it increases at an intermediate rate and then at a more rapid rate.
For this fifth power polynomial velocity case, the coupling between the random motion in the transverse direction and the flow velocity gradient significantly alters not only the variance of particle displacement in the streamwise direction but also the particle mean displacement, which is also observed in the case of plane-Poiseuille flow. Suppose a group of particles is released freely and independently into the hyperbolic tangent flow at the point of origin. The evolution of the mean displacement of all the particles compared to $v_f(0) t$ follows the blue curve in Figure \ref{InHT}. Particles will disperse in the streamwise direction with a standard deviation compared to $2D^*t$ following the square root of the orange curve in Figure \ref{InHT}.

\subsection{Discussion}  \label{disc}

The exact particle MSD has been derived for the Couette flow of a first-order polynomial velocity profile, plane-Poiseuille flow of a second-order polynomial velocity profile, and fifth-order polynomial expansion of a hyperbolic tangent flow. The anomalous diffusion in these three cases implies that the flow velocity gradient along the transverse direction plays a significant role in the diffusion of Brownian particles in the streamwise direction.

\begin{figure}[htbp]

\centering
\includegraphics[width=10cm]{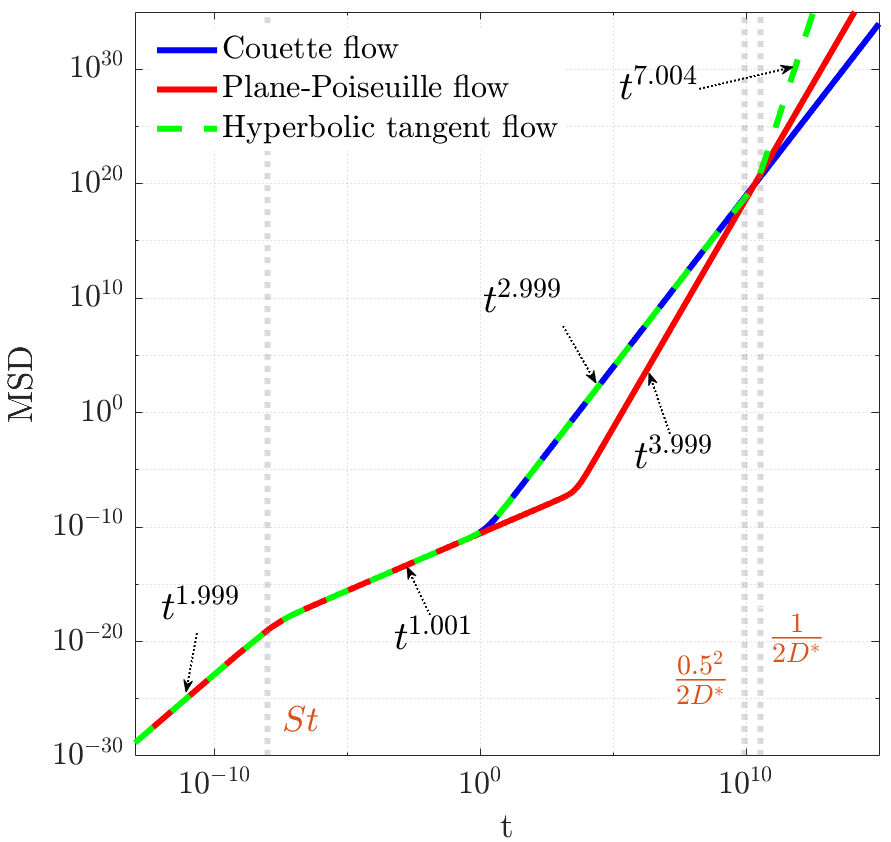}
\caption{Comparison of the particle MSD in Couette flow, plane-Poiseuille flow, and Hyperbolic tangent flow at all time regions for $St = 10^{-8}$ when particle initially located in the centerline in the transverse direction.}
\label{com3}

\end{figure}

The particle MSD in the streamwise direction calculated above includes advection effects (expressed as $v_f(0) t$), particle pure diffusion due to the random motion in the streamwise direction (expressed as $2D^*t$ in the variance of particle displacement), and the coupling between flow velocity gradient and particle diffusion in the transverse direction (expressed both in the mean and variance of particle displacement). For clarity, we subtract the advection effect due to fluid motion from particle displacement. The exact particle MSD, caused by pure diffusion and the coupling between flow velocity gradient and particle diffusion in the transverse direction, are presented in Figure \ref{com3} for the Couette, plane-Poiseuille, and Hyperbolic tangent flows.

From Figure \ref{com3}, we observe that for all three cases, the particle MSD is a function of $t^{1.999}$ at short-time scales ($t \ll St$), which follows the short-time analysis for which $u_{p x}^2(0) t^2$ dominates. 
Starting from $t \sim \mathcal{O}(St)$, the particle MSD changes into a function of $t^{1.001}$ at $St \ll t \ll \mathcal{O}(1)$, and consequently particle pure diffusion dominates in the streamwise direction. This can also be confirmed by the variance in equations (\ref{msd-linear-long}) and (\ref{msd-para-long}) where $\sigma_{s_p}^2 \sim 2 D^* t$ at $St \ll t \ll \mathcal{O}(1)$ for $c_n \sim \mathcal{O}(1)$ and $y_p(0) \sim \mathcal{O}(1)$. When the normalized time $t$ is much larger than $\mathcal{O}(1)$, the coupling between the flow velocity gradient and the particle diffusion in the transverse direction starts to affect the results. For the cases of Couette flow (plane-Poiseuille flow), the particle MSD is a function of $t^{2.999}$ ($t^{3.999}$), which may be confirmed by the variance in equation (\ref{msd-linear-long}) (equation (\ref{msd-para-long})) for which the $t^3$-term ($t^4$-term) dominates in the variance when $t \gg \mathcal{O}(1) \gg St$. As for the polynomial expansion of hyperbolic tangent flow, the temporal evolution of the particle MSD is first a function of $t^{2.999}$ and then shifts into $t^{7.004}$ when $t \gg \frac{1}{2D^*}$.

In the previous analysis, we consider the parallel shear flow as unbounded in the transverse direction. However, realistically,  boundaries exist in the transverse direction of a flow. Hence, an observation time interval for particle diffusion may be defined from $t=0$ when particles are initially introduced at their initial position and the time at which the particles reach the boundaries in the transverse direction. Since particles exhibit simple diffusion in the transverse direction, the time interval $T_v$, at which the particle reaches the boundaries in the streamwise direction, can be roughly estimated by
\begin{equation}
  T_v = \frac{L_{BP}^2}{2 D^*} ~,
\end{equation}
with $L_{BP}$ the minimum distance between the particle and the boundaries. $L_{BP}$ may be expressed as 
\begin{equation}
    L_{BP} = \min \left( \lvert y_{B1} - y_p(0) \rvert, \lvert y_{B2} - y_p(0) \rvert \right)~, 
\end{equation}
where $y_{B1}$ and $y_{B2}$ are dimensionless coordinate of the flow boundaries. For example, when a particle is initially located at $y_p(0) = 0.7$ in the dimensionless Couette flow, the observation time of particle diffusion should be much less than $T_v = 0.3^2/(2D^*)$. When particles are initially located in the middle between the two boundaries, the observation time is maximum. For Couette flow, the maximum observation time is $0.5^2/(2D^*)$ since the dimensionless coordinate of the flow boundaries are $y_{B1} = 0$ and $y_{B2} = 1$. For plane-Poiseuille and hyperbolic tangent flows, the maximum observation time is $1/(2D^*)$ since the dimensionless coordinate of the flow boundaries are $y_{B1} = -1$ and $y_{B2} = 1$. Figure \ref{com3} presents the change of the particle MSD in the streamwise direction in the three studied cases and demonstrates how the estimated limit of corresponding observation time correlates to the shift in MSD.

The differences in the three studied cases stem from the magnitude and gradient of the shear flow velocity. 
The magnitude of the flow velocity determines the particle mean displacement, which is already subtracted here. However, the velocity gradient alters the particle diffusion (shown in the variance of particle displacement) through the random motion in the transverse direction. 
Figure \ref{com3} indicates that, within the maximum observation time, particle MSD behaves similarly in the Couette flow and the polynomial expansion of hyperbolic tangent flow. It implies that the flow velocity gradient close to the centreline in the Couette flow is similar to that of the hyperbolic tangent flow around the centerline between $y=-1$ and $y=1$, presented in Figure \ref{fig-linear-parab}. However, the flow velocity gradient at the centerline in the plane-Poiseuille flow is different from that of the Couette flow, resulting in the distinguishable MSD shown in Figure \ref{com3}.

\begin{figure}[ht]
\centering
\includegraphics[width=10cm]{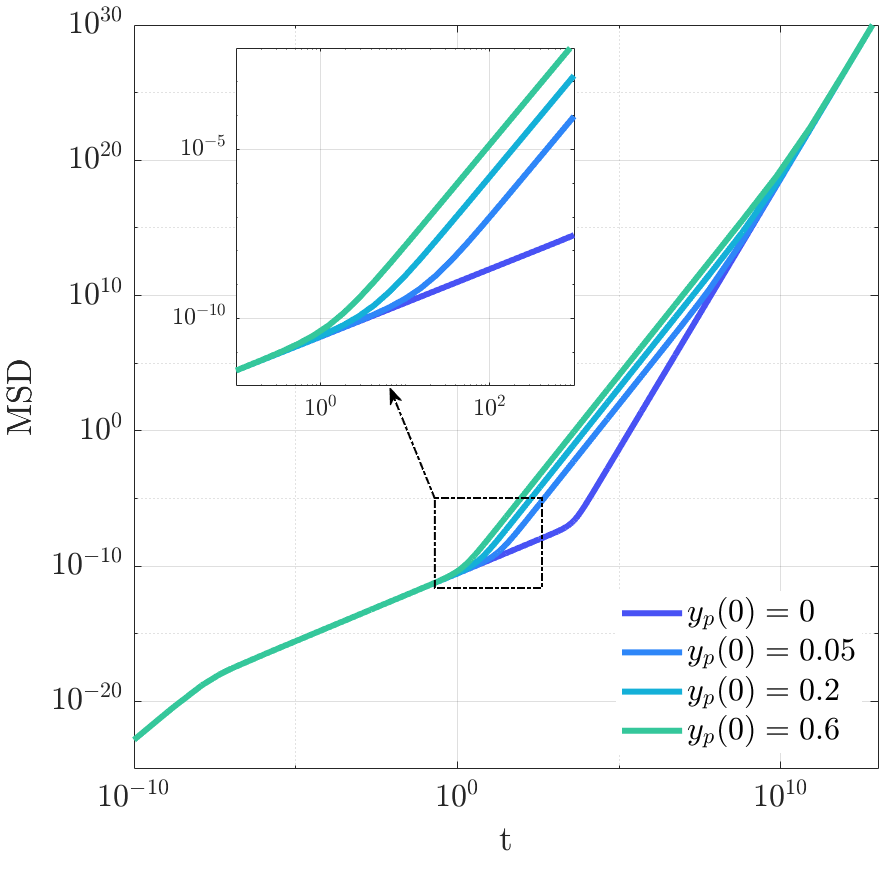}
\caption{Velocity gradient effects on the variance of particle displacement in plane-Poiseuille flow for $St = 10^{-8}$.}
\label{GradientEffects}
\end{figure}

\begin{figure}[ht]
\centering
\includegraphics[width=10cm]{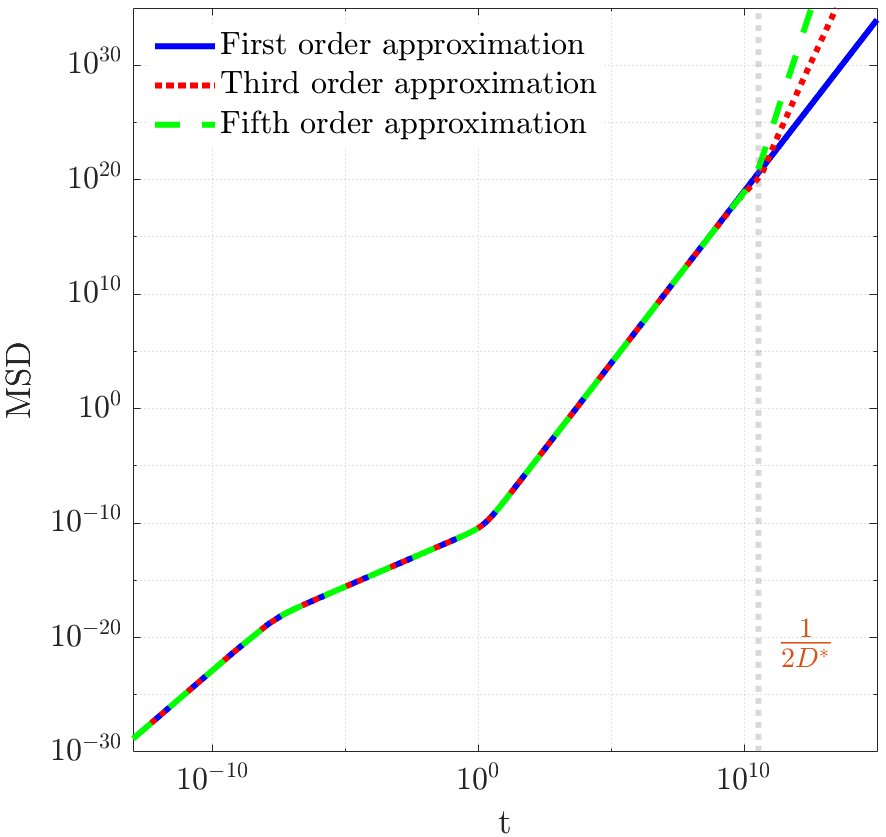}
\caption{Effects of the number of polynomial expansion terms in hyperbolic tangent flow with $y_p(0) = 0$ and $St =10^{-8}$.}
\label{Numffects}
\end{figure}

Moreover, in equation (\ref{msd-linear-long}), we observe the dependence of the variance of particle displacement on $c_1$, which is the coefficient of the first and the only power of $y$ in the velocity profile. However, in the case of plane-Poiseuille flow, the variance in equation (\ref{msd-para-long}) depends not only on $c_2$ -- the coefficient of the second and the only power of $y$ in the velocity profile -- but also on particle initial position $y_p(0)$.
The reason is that the velocity gradient is constant everywhere in the Couette flow but varies along the transverse coordinate in the plane-Poiseuille flow. Figure \ref{GradientEffects} illustrates the effects of the velocity gradient on particle diffusion when particles are initially located at a position with different velocity gradients. Since the flow profile is symmetric about the centerline, in Figure \ref{GradientEffects}, we only show the initial position effect from the upper side. In plane-Poiseuille flow, the further away from the centerline, the higher the velocity gradient is. Figure \ref{GradientEffects} shows that the higher the velocity gradient at the particle's initial position, the higher diffusion (MSD) particles obtain and the earlier the particle MSD shifts to a function of high-order temporal evolution.

As for the number of polynomial expansion terms of a flow, clearly, the more terms are taken, the more accurately the particle MSD will be obtained. At the same time, the calculation of the particle MSD including more expansion terms will be demanding. However, for the studied hyperbolic tangent flow, Figure \ref{com3} shows that within the valid observation time, the first power of $y$ in equation (\ref{htpfun}) dominates when particles are initially located at the centerline. Figure \ref{Numffects} shows that the first, third, and fifth-order polynomial expansion of hyperbolic tangent flow characterizes the variance well. For this case, we deduce that other terms may affect particle mean displacement but have little influence on the variance of particle displacement. For any polynomial expansion of the flow, one may conduct the same analysis to determine how many terms should be taken.

\section{Concluding Remarks} \label{conclu}

In this study, we mathematically derived a new formulation for Brownian particle diffusion in two-dimensional parallel shear flows that may be described by a polynomial velocity profile.
Based on the Langevin equation, stochastic calculus was employed to resolve the generalized formulation, assuming the product of the random force and the time interval follows the increment of a Wiener process during this time interval.

Using this approach, we demonstrate that the particle MSD is the same as in a stationary medium in the limit of short-time scales, as expected. 
For long-time scales relative to $St$ in general polynomial laminar flows, we find that the order of $t$ in particle diffusion is $n+2$ where $n$ is the polynomial order of the transverse coordinate in the flow velocity profile.
Our method was validated for two cases: Couette flow of a linear velocity profile and plane-Poiseuille flow of a parabolic velocity profile, converging to the results previously obtained using differential calculus.
This stochastic solution method can also apply to any parallel shear flow, of which the velocity profile may be expanded into a polynomial series. As an example, a fifth-order polynomial expansion for a hyperbolic tangent shear flow was discussed and analyzed.

The method presented here also bridges the particle MSD at short and long-time scales and resolves the exact particle MSD for all-time regions, which may contribute to a more accurate prediction of particle diffusion with higher spatiotemporal resolution. 
The three studied cases show significant effects of the shear flow velocity profile on particle diffusion in the streamwise direction. Moreover, three distinct regions are revealed for the particle MSD along the timeline, demonstrating different physical mechanisms dominating the particle diffusion at different observation times.
Thus, our study suggests a new generalized formulation for the diffusion of Brownian particles that may be applicable to any two-dimensional parallel laminar flows at all time scales.

\section*{Acknowledgments} This research was supported by the ISRAEL SCIENCE FOUNDATION (grant No. 1762/20).

\section*{Appendix}
\label{appen}

For the time integral of powers of Brownian motion, by stochastic Fubini theorem and It$\hat{o}$'s lemma, 
 \begin{equation*}
    \begin{aligned} 
    \int_0^t  \mathbb{W}_y (u) d u  &=  \int_0^t  \int_0^u d\mathbb{W}_y (s) d u  =  \int_0^t  \int_{s}^t d u  d\mathbb{W}_y (s) =  \int_0^t  (t -s)  d\mathbb{W}_y (s)  ~;
    \end{aligned}
\end{equation*}
\begin{equation*}
    \begin{aligned} 
     \int_0^t  \mathbb{W}_y^2 (u) d u & = \int_0^t \int_0^u d\mathbb{W}_y^2 (s) d u =  \int_0^t \int_0^u 2 \mathbb{W}_y(s) d\mathbb{W}_y(s) d u +\int_0^t \int_0^u  d s d u \\&= \int_0^t \int_s^t 2 \mathbb{W}_y(s) d u d\mathbb{W}_y (s) + \frac{t^2}{2}  = \int_0^t 2 (t-s)\mathbb{W}_y(s)  d\mathbb{W}_y(s) + \frac{t^2}{2}     ~;
    \end{aligned}
\end{equation*}
\begin{equation*}
    \begin{aligned} 
     \int_0^t  \mathbb{W}_y^3 (u) d u & = \int_0^t \int_0^u d\mathbb{W}_y^3 (s) d u =  \int_0^t \int_0^u 3 \mathbb{W}_y^2(s) d\mathbb{W}_y(s) d u +\int_0^t \int_0^u 3 \mathbb{W}_y(s) d s d u \\&=  \int_0^t \int_s^t 3 \mathbb{W}_y^2(s) d u d\mathbb{W}_y(s) +\int_0^t \int_s^t 3 \mathbb{W}_y(s) d u d s \\&=  \int_0^t 3(t-s)\mathbb{W}_y^2(s) d\mathbb{W}_y(s) +\int_0^t 3 (t-s)\mathbb{W}_y(s) d s \\&=  \int_0^t 3(t-s)\mathbb{W}_y^2(s) d\mathbb{W}_y(s) + \int_0^t 3 (t-s)\int_0^s d\mathbb{W}_y(u) d s  \\&=  \int_0^t 3(t-s)\mathbb{W}_y^2(s) d\mathbb{W}_y(s) + \int_0^t\int_s^t 3 (t-s) d s d\mathbb{W}_y(u) \\&=  \int_0^t 3(t-s)\mathbb{W}_y^2(s) d\mathbb{W}_y(s) + \int_0^t \frac{3}{2} (t-s)^2 d\mathbb{W}_y(s)    ~;
    \end{aligned}
\end{equation*}
\begin{equation*}
    \begin{aligned} 
     \int_0^t  \mathbb{W}_y^4 (u) d u & = \int_0^t \int_0^u d\mathbb{W}_y^4 (s) d u =  \int_0^t \int_0^u 4 \mathbb{W}_y^3(s) d\mathbb{W}_y(s) d u +\int_0^t \int_0^u 6 \mathbb{W}_y^2(s) d s d u \\&=  \int_0^t 4(t-s)\mathbb{W}_y^3(s) d\mathbb{W}_y(s) +\int_0^t 6 (t-s)\mathbb{W}_y^2(s) d s \\&=  \int_0^t 4(t-s)\mathbb{W}_y^3(s) d\mathbb{W}_y(s) + \int_0^t\int_0^s 12 (t-s) \mathbb{W}_y(u) d\mathbb{W}_y(u) d s + \int_0^t\int_0^s  6 (t-s) d u d s \\&=  \int_0^t 4(t-s)\mathbb{W}_y^3(s) d\mathbb{W}_y(s) + \int_0^t\int_u^t 12 (t-s) \mathbb{W}_y(u) d s d\mathbb{W}_y(u) + t^3 \\&=  \int_0^t 4(t-s)\mathbb{W}_y^3(s) d\mathbb{W}_y(s) + \int_0^t 6 (t-s)^2 \mathbb{W}_y(s) d\mathbb{W}_y(s) + t^3  ~;
\end{aligned}
\end{equation*}
~~~~~~~~~~~~~~~~~~~~~~~~~~~~~~~~~~\vdots

When $n$ is odd, 
\begin{equation*}
    \begin{aligned}
     \int_0^t  \mathbb{W}_y^n (u) d u & = \int_0^t \int_0^u d\mathbb{W}_y^n (s) d u =  \sum_{l=1}^{ \frac{n+1}{2}} \frac{n!}{2^{l-1}(n-2l+1)!}\int_0^t  \frac{(t-s)^l}{l!}\mathbb{W}_y^{n-(2l-1)}(s) d \mathbb{W}_y(s)   ~.
    \end{aligned}
\end{equation*}

And when $n$ is even,
\begin{small}
\begin{equation*}
    \begin{aligned}
     \int_0^t  \mathbb{W}_y^n (u) d u & = \int_0^t \int_0^u d\mathbb{W}_y^n (s) d u =  \sum_{l=1}^{ \frac{n}{2}} \frac{n!}{2^{l-1}(n-2l+1)!}\int_0^t  \frac{(t-s)^l}{l!}\mathbb{W}_y^{n-(2l-1)}(s) d \mathbb{W}_y(s)  + (n-1)!!\frac{2}{n+2}t^{\frac{n+2}{2}}  ~.
    \end{aligned}
\end{equation*}
\end{small}

\bibliography{ref.bib}

\end{document}